\documentclass[12pt,a4paper]{article}
\usepackage{units}
\usepackage{hyperref}
\usepackage{amsmath,amssymb}
\usepackage{units}
\usepackage{graphicx}
\input ifpdf.sty
\ifpdf
\fi
\newcommand{\gew}{g_5}
\newcommand{\gewt}{\tilde{g_5}}
\newcommand{\gs}{g_{5S}}
\newcommand{\mslash}[1]{#1\hspace{-1.5ex}\slash\hspace{0.3ex}}
\newcommand{\etmiss}{\mslash{E}_T}
\newcommand{\ptmiss}{\mslash{P}_T}
\newcommand{\pt}{P_T}
\newcommand{\pointAc}{P1}
\newcommand{\pointAa}{P2}
\newcommand{\pointAb}{P3}
\newcommand{\pointB}{P4}
\newcommand{\pointD}{P5}
\newcommand{\pointDb}{P6}

\allowdisplaybreaks

\title{%
  Production of Dark~Matter in Warped~Higgsless~Models
  with Composite~Sector~Supersymmetry}
\author{%
  Alexander Knochel%
    \thanks{e-mail: \texttt{aknochel@physik.uni-wuerzburg.de}}
    ${}^{a,b}$\\
  Thorsten Ohl%
    \thanks{e-mail: \texttt{ohl@physik.uni-wuerzburg.de}}
    ${}^{a}$\\
  \hfil\\
    ${}^a$Institut f\"ur Theoretische Physik und Astrophysik\\
    Universit\"at W\"urzburg\\
    Am Hubland, 97074 W\"urzburg, Germany\\
  \hfil\\
    ${}^b$Albert-Ludwigs Universit\"at Freiburg\\
    Institut f\"ur Physik\\
    Hermann-Herder-Str. 3, 79104 Freiburg, Germany}

\begin{document}

\maketitle
\begin{abstract}
  We explore the LHC phenomenology of an extension of warped higgsless
  models.  The model is supersymmetric in the bulk and on the IR brane
  as introduced in~\cite{Knochel:2008ks}, corresponding to an
  emergence of a supersymmetric spectrum in the composite sector of
  the higgsless model. In particular, the lightest neutralino is
  rendered stable by an~$R$ parity and serves as a realistic cold dark
  matter candidate. The observation of missing energy signals at the
  LHC from LSP and NLSP production in association with third
  generation quarks is discussed based on simulations using WHIZARD.
\end{abstract}

\section{Introduction}
Two of the most prominent issues to be addressed at the Large Hadron
Collider (LHC) are the origin of electroweak symmetry breaking~(EWSB)
and the nature of dark matter (DM) which, according to recent results
of the WMAP experiment, makes up roughly 20\%{} of the energy density
of the observable universe.  The Standard Model~(SM) of particle
physics, when extended to include massive neutrinos, is extremely
successful at describing the observations made at past and present
experiments, including many nontrivial precision tests.  Despite these
achievements, it fails to provide a candidate for cold DM in the form
of a stable, weakly interacting particle species, but at the same time
predicts a scalar Higgs particle which so far has eluded detection at
the LEP and Tevatron experiments. In addition, the minimal Higgs
mechanism is theoretically unappealing due to the severe fine tuning
required by quadratically divergent radiative corrections to
the Higgs potential.  It is probably not a coincidence that all of the
problems mentioned above, the observed DM relic density $\Omega h^2
\approx 0.1$ and the shortcomings of the standard EWSB scenario,
strongly point to a modification and/or extension of the SM at the TeV
scale which is thus within reach of the LHC experiments. We take the
apparent absence of the SM Higgs boson below $\approx 115$ GeV and the
cosmological evidence for weakly interacting stable particles at or
below the TeV scale as a motivation to consider an alternative
scenario which relates both issues, namely a supersymmetric
formulation of higgsless models in a warped extra dimension in which
the properties of the gaugino dark matter candidate are closely tied
to the mechanism which produces the electroweak gauge boson masses.
We have introduced the model setup in \cite{Knochel:2008ks} including
a discussion of the virtues of the lightest neutralino modes as a cold
dark matter (CDM) candidate.

This paper is organized as follows: In section~\ref{sect_model} we give a
short review of the model including the superfield content, 
the boundary conditions imposed to obtain a realistic
light spectrum, and some key properties. We discuss the parameters
which have an important impact on LSP phenomenology at the LHC, and
define a set of benchmark points in parameter space.  In
section~\ref{sect_lhc} we discuss promising channels for an observation of
missing energy from LSP pair production. Our conclusions are based on
the simulation of four parton final states at the LHC with
WHIZARD~\cite{WHIZARD}. We propose sets of kinematical cuts and present the
corresponding results for a range of model parameters. In
section~\ref{sect_conclusions} we summarize our results.

\section{The Model}
\label{sect_model}
\subsection{General Setup}

The model~\cite{Knochel:2008ks} on which we base the following
discussions is an extension of the warped
higgsless model proposed in~\cite{Csaki/etal}. The 5D gauge group is
\begin{equation}
G_{bulk}=SU(3)_C\times SU(2)_L \times SU(2)_R \times U(1)_{B-L}\,,
\end{equation}
with the corresponding 5D gauge coupling constants $\gs$, $g_{5L}=g_{5R}=\gew$ and $\gewt$. 
The full symmetry is conserved in the bulk and broken by boundary conditions to
\begin{equation}G_{bulk}\rightarrow
\begin{cases}
 G_{SM}=SU(3)_C\times SU(2)_L\times U(1)_Y     &\quad\text{on the UV brane}\\
 \hphantom{G_{SM}=}
        SU(3)_C\times SU(2)_D\times U(1)_{B-L} &\quad\text{on the IR brane}
\end{cases}
\end{equation}
and thus only $SU(3)_C\times U(1)_{EM}$ survives as an unbroken
subgroup.  Matter hypermultiplet masses are generated on the IR brane
as usual where they are allowed by the $SU(2)_D$ symmetry, while the
isospin doublets are split by Majorana masses and localized kinetic
terms on the UV brane where $SU(2)_R$ is broken.

Since the smallest spinors in 5D are of Dirac type, there are four
supercharges in 5D $\mathcal{N}=1$ SUSY, which can be related to 4D
$\mathcal{N}=2$ SUSY. However, half of the symmetries are broken by
the background, leaving us with usual $\mathcal{N}=1$ SUSY after
Kaluza-Klein expansion. Nevertheless, we assume that full 5D
hypermultiplets with corresponding interactions are present in the
bulk of the extra dimension in such a way that full 5D SUSY spectra
appear at energies $E>\Lambda_{IR}$. The corresponding action of
hypermultiplets coupled to 5D Super Yang-Mills theory broken by a
warped background can be expressed by ordinary $\mathcal{N}=1$ chiral-
and vector superfields~\cite{HEIDISUSY}.  The resulting superfield
content with representations and quantum numbers is summarized in
table \ref{fig_sfields}.  Each hypermultiplet has a bulk mass
\begin{equation}
M_i=c_i k\,,
\end{equation}
parameterized in units of the Randall-Sundrum curvature $k$ which acts
as a localization parameter for the entire multiplet.
\begin{table}
\begin{center}
\begin{tabular}{llll}
Field& Representations & Field & Representations\\
\hline
 $V^{Ca}$, $\chi^{Ca}$ &$SU(3)_C$ gauge multiplet &
 $V^{Li}$, $\chi^{Li}$ &$SU(2)_L$ gauge multiplet \\
 $V^{Ri}$, $\chi^{Ri}$ &$SU(2)_R$ gauge multiplet &
 $V^X$, $\chi^X$ & $U(1)_{B-L}$ gauge multiplet \\[2ex]
 $H^{L}_{l,g}$&$(\mathbf{1},\mathbf{2},\mathbf{1})_{-1}$ &
 $H^{R}_{l,g}$&$(\mathbf{1},\mathbf{1},\mathbf{2})_{-1}$ \\
 $H^{Lc}_{l,g}$&$(\mathbf{1},\mathbf{\overline{2}},\mathbf{1})_{1}$ &
 $H^{Rc}_{l,g}$&$(\mathbf{1},\mathbf{1},\mathbf{\overline{2}})_{1}$\\[2ex]
 $H^{L}_{q,g}$&$(\mathbf{3},\mathbf{2},\mathbf{1})_{1/3}$ &
 $H^{R}_{q,g}$&$(\mathbf{3},\mathbf{1},\mathbf{2})_{1/3}$  \\
 $H^{Lc}_{q,g}$&$(\mathbf{\overline{3}},\mathbf{\overline{2}},\mathbf{1})_{-1/3}$ &
 $H^{Rc}_{q,g}$&$(\mathbf{\overline{3}},\mathbf{1},\mathbf{\overline{2}})_{-1/3}$
\end{tabular}
\caption{\label{fig_sfields}%
  The superfield content of the model and corresponding
  representations and quantum numbers with respect to the bulk gauge
  group $G_{bulk}=SU(3)_C\times SU(2)_L \times SU(2)_R\times
  U(1)_{B-L}$. Here, $V$ denotes vector superfields, $\chi$ denotes
  chiral superfields in the adjoint as they appear in $\mathcal{N}=2$
  vector multiplets, and $H$ the denotes the chiral superfield
  component of the matter hypermultiplets. $g=1\dots 3$ denotes the
  generation.}
\end{center}
\end{table}

To define the full model, one needs to specify the boundary conditions
and brane action. The IR brane (i.\,e.~$y=\pi$) boundary
conditions are a straightforward generalization of the
nonsupersymmetric boundary conditions \cite{Knochel:2008ks}. In
contrast, SUSY is broken on the UV brane, which means that boundary
conditions can differ within 4D multiplets.  We assume a localized
kinetic term for the $SU(2)_R$ transforming fields on the UV brane
controlled by a parameter $\rho$, which yields effective boundary
conditions after Kaluza-Klein expansion
\begin{equation}
f_q^R(0)=m \rho^2 f_{q^c}^{R}(0)\,,
\end{equation}
while neutrinos can receive a UV localized Majorana mass~$\mu_r$,
leading to effective boundary conditions,
\begin{equation}
\psi_\nu^R(0)=\frac{\mu_r}{k} \psi_\nu^{Rc}(0)\,.
\end{equation}
The physical scalars $h_f$, $h^c_f$, $\Sigma^i$ get universal Dirichlet conditions
\begin{equation}
h_f(0)=h^c_f(0)=\Sigma^i(0)=0
\end{equation}
and the gluinos receive boundary conditions that are ``twisted'' with
respect to the gluons
\begin{equation}
\lambda_1^C(0)=0\,.
\end{equation}
For the electroweak gauginos there are several possibilities. The
neutral sector consists of three 5D gauginos, $\lambda^{L3}_i$,
$\lambda^{R3}_i$, $\lambda^X_i$, while the charged sector is spanned
by $\lambda_i^{L\pm}$ and $\lambda_i^{R\pm}$. The boundary conditions
of the charged and neutral sectors are linked by the $SU(2)_L$
symmetry. The neutralino boundary conditions can be classified by the
number of twisted/untwisted conditions imposed. Furthermore, the
fields $\lambda^X$ and $\lambda^{R3}$ can mix on the UV
brane. However, only two choices remove all massless gaugino modes
automatically from the spectrum, both with two twisted boundary
conditions in the neutral sector. The first one is
\begin{subequations}
\begin{equation}
\label{eq:scenario-1}
  \lambda_2^{L}= \lambda_1^{R}= \lambda_1^{X}=0\,,
\end{equation}
while in the second
\begin{equation}
\label{eq:scenario-2}
\begin{aligned}
  \lambda^{L}_1
   &= \lambda^{R12}_2 =0 \\
      \cos(\theta_N) \lambda^X_1+\sin(\theta_N) \lambda^{R3}_1
   &= \cos(\theta_N) \lambda^{R3}_2-\sin(\theta_N) \lambda^X_2
    = 0\,.
\end{aligned}
\end{equation}
\end{subequations}
Since in the former, charginos and neutralinos are degenerate with the
$W$ bosons and neutralino annihilation cross sections are too large,
we choose the latter \cite{Knochel:2008ks}. This completes the list of
boundary conditions.

\subsection{Enhanced Cross Sections from Almost Delocalized Quarks}
While the parameter space of the model is still rather large, there
are several reasonable assumptions that one can make. First of all, at
tree level, we impose degeneracy of the pairs of electroweak gaugino
modes which will be lifted only at the loop level (i.\,e.~no Majorana
masses on the UV brane). Furthermore, the splitting of the $W$ and
$\chi^+$ raises the KK scale and is therefore assumed to be as small as
possible in order to achieve unitarization (i.\,e.~with the lightest
charginos just above the experimental exclusion range). The neutralino
mixing angle $\theta_N$ is then fixed by the relic density. The
localization of the light quarks is largely constrained by the $S$
parameter to be close to $c_L = 0.5$, which also suppresses the
coupling to the heavy resonances. The third generation is naturally IR
localized to generate the heavy top.

While this basically fixes the properties that are relevant to the LSP
production processes which we will discuss later, there remains some
freedom in the localization parameters which strongly impact LHC
phenomenology. Let us therefore discuss them in more detail.  Exactly
delocalized light quarks ($c_L=-c_R=1/2$) have vanishing couplings to
KK gluons, while a small deviation from delocalization introduces
nonzero couplings. At the same time, localized kinetic terms for the
quarks which are used to split the isospin doublets, introduce a localization
effect on the UV brane which also shifts these effective couplings to
a nonzero value.

\begin{figure}
\begin{center}
\includegraphics[width=9cm]{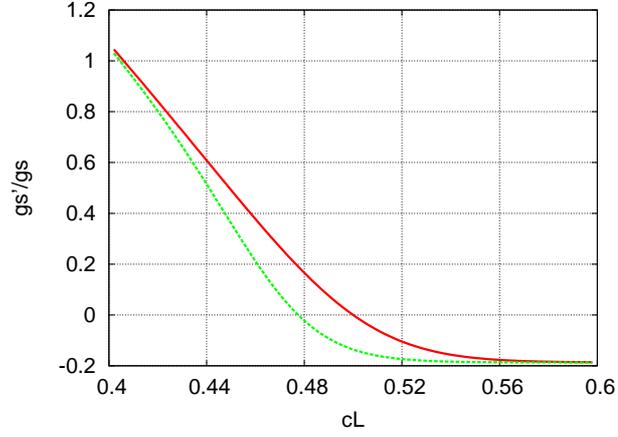}
\caption{The coupling of a massless quark to the first gluon resonance
  as a function of the bulk mass. The upper line is for $\rho=0$,
  while the lower one is for $\rho^2 k=100$, which corresponds roughly
  to the $u-d$ splitting at $c_L=-c_R=1/2$. \label{fig_quarklocal}}
\end{center}
\end{figure}

The importance of KK gluon intermediate states in our study of LSP
production at the LHC depends on the exact choices of localization
parameters.  Our minimal implementation of the third generation,
though not addressing the $Zbb$ problem~\cite{Zbb}, provides a simple
way to study the phenomenology of the $t$ and $b$ in LSP production
for different scenarios from strongly IR localized fields to the
almost delocalized case. The coupling strength of a (righthanded) quark zero
mode
\begin{equation}
f_{\psi}=N z^{2+c_R}
\end{equation}
to the KK gluon,
\begin{equation}
\frac{g_s'}{g_s \sqrt{\pi R}}=\langle f_R f_R f_g' \rangle = \frac{\int \! \left[(zk)^{-4}+\rho_q^2\delta(z-k^{-1})\right] dz \, z^{4+2c_R}  f_g'}{\rho_q^2 k^{-2 c_R-4}+  \int \! (zk)^{-4} dz \, z^{4+2c_R} }
\end{equation}
is shown in figure~\ref{fig_quarklocal} as a function of the bulk mass
$c_L=-c_R$. As expected, the introduction of a UV localized kinetic
term for the quarks shifts the effective localization. While the
localization of the third generation lets the mass of the first quark
KK modes vary between extremely light (for almost delocalized third
generation fields requiring large IR Dirac masses) and heavy
($\approx 3 k e^{- R k \pi}$), the masses of the lightest $\tilde{t}$
and $\tilde{b}$ modes stay below $2 k e^{- R k \pi}\approx 1100$ GeV
and can thus be pair produced at LHC energies with appreciable cross
sections.  We thus define in table~\ref{tab:Pn} five points in parameter
space which represent extreme and intermediate cases of quark
localization,

\begin{table}
\begin{center}
\begin{tabular}{c|c|c|c|c|c|c}
Bulk Mass&\pointAc{}&\pointAa{}&\pointAb{}&\pointB{}&\pointD{}&\pointDb{}\\
\hline\hline $c_{L1,2}$ &0.48&0.48&1/2&0.48&0.48&0.48 \\
\hline $c_{R1,2}$ &-0.48&-0.48&-1/2&-0.48&-0.48&-0.48 \\
\hline $c_{L3}$ &1/3&0.4&0.4& 0.2&0.2&0.4\\
\hline $c_{R3}$ &-0.4&-1/3&-1/3& -0.2&-0.4&-0.2\\
\end{tabular}
\end{center}
\caption{\label{tab:Pn}%
  Definitions of points in bulk mass parameter
  space of the first and second ($c_{1,2}$) and third ($c_3$)
  generation of quarks.}
\end{table}
The points \pointAc{} and \pointAa{} feature slightly IR localized
light quarks corresponding to vanishing tree level contributions to
$S$, while having small but nonzero couplings to KK gluons. In
\pointAc{}, the bulk masses of $SU(2)_L$ and $SU(2)_R$ transforming
hypermultiplets are exchanged relative to \pointAa{}. \pointAb{} is a
variation of \pointAa{} with exactly delocalized light lefthanded
quarks.  \pointB{} features rather strongly IR localized third
generation hypermultiplets, while the points \pointD{} and \pointDb{}
correspond to a more asymmetric third generation localization.

\begin{figure}
\begin{center}
\includegraphics[width=9cm]{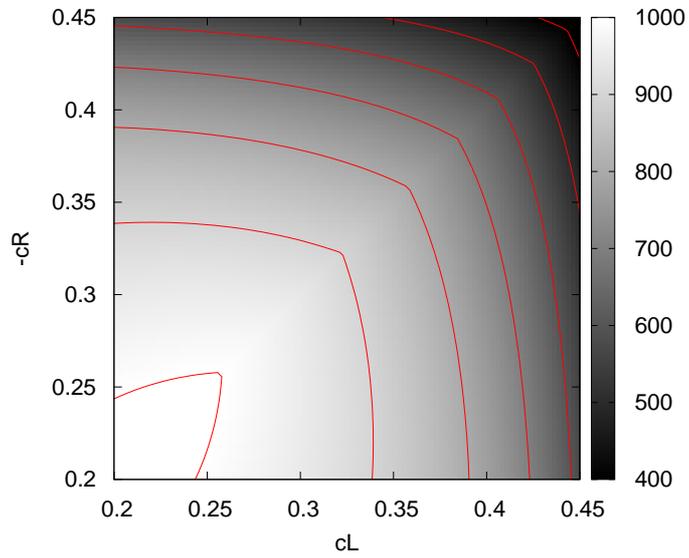}
\caption{The mass of the lightest stop resonance in GeV as a function
  of the third generation bulk masses for $\Lambda_{IR}\approx 600$
  GeV.\label{fig_stopmass}}
\end{center}
\end{figure}
\subsection{The Particle Spectrum}
\label{sect_spect}
The spectrum of new particles in the LHC range is given by the KK
resonances and their $\mathcal{N}=2$ superpartners residing at $600
\dots 1500$ GeV, some of which are very broad (e.g. the color charged
resonances $\tilde{g}, g'$), while the $f'$, $\tilde{f}$ can be
comparatively narrow with $\Gamma<100$ GeV. Due to boundary
conditions, there are no sfermion or gluino zero modes, making
otherwise unprotected UV localized light scalars unnatural.  The
relevant two body decay modes of the light $\tilde{q}$ are into
$\chi^{\pm0} q$. However, as soon as any resonance is heavy enough to
allow KK fermions or KK gauginos in the final state those dominate due
to the strongly enhanced couplings. The first KK fermions have strong
electroweak two body decays $f\rightarrow W/Z, f$, while the KK
charginos and neutralinos themselves mainly decay into $t\tilde{t}$,
$b\tilde{b}$, gluinos in addition to $t'\tilde{t}$, $b'\tilde{b}$, both
with enhanced couplings due to IR localization of the third
generation. For the same reason, the gluon decays to
$t\overline{t}$, $t'\overline{t}$ etc. with total widths varying
between $\approx 300 \dots 500$ GeV, while the scalar gluon coupling
is very sensitive to quark localization and vanishes in the LR
symmetric case, resulting in widths ranging from $\Gamma\approx
10\dots 100$ GeV in which the $t'\bar t$ final state dominates.  The
masses and widths of the relevant color charged resonances are summarized in
table~\ref{tabspectrum}.

\begin{table}
\begin{center}
\begin{tabular}{c|cccccc}
Pt.& \pointAa{}& \pointAb{}& \pointAc{}& \pointB{}& \pointD{}& \pointDb{} \\ \hline
 $t^1$ & 970(30) & 970(30) & 970(40) & 1480(90) & 1180(60) & 1180(60) \\ \hline
 $t^2$ & 2140(460) & 2140(460) & 2140(480) & 2050(100) & 2120(340) & 2120(300)  \\ \hline
 $\tilde{t}^1$ & 880(40) & 880(40) & 800(20) & 1100(40) & 860(20) & 1080(110) \\ \hline
 $\tilde{t}^{1c}$ & 800(80) & 800(80) & 880(130) & 1100(140)& 1080(230) & 860(80)   \\ \hline
 $\tilde{t}^{2}$ &  2040(450) &  2040(450) &  1970(320) & 1850(180) & 1840(120) & 2020(420)  \\ \hline
 $\tilde{t}^{c2}$ & 1970(350)  &  1970(350)  & 2040(490) & 1850(340) & 2020(490) & 1840(160)
\end{tabular}\\[3ex]
\begin{tabular}{c|cccccc}
Pt.& \pointAa{}& \pointAb{}& \pointAc{}& \pointB{}& \pointD{}& \pointDb{} \\ \hline
  $b^1$  &  800(40) &  800(40) &  880(50)  &  1100(60)&  1080(80)&  860(50)  \\ \hline
 $b^2$ & 1970(470) & 1970(470) & 2040(480)&  1850(150)&  2020(340)& 1840(320) \\ \hline
 $\tilde{b}^{1}$  & 880(80)& 880(80)& 800(40)& 1100(80)& 860(40) & 1080(160) \\ \hline
 $\tilde{b}^{c1}$ & 800(70) & 800(70) & 880(120)& 1100(130)& 1080(220)& 860(70) \\ \hline
 $\tilde{b}^2$  & 2040(280)& 2040(280)& 1970(270)& 1850(90)& 1840(110) & 2020(170) \\ \hline
 $\tilde{b}^{c2}$  & 1970(380)  & 1970(380)  & 2040(540)  & 1850(340)  & 2020(560)  & 1840(190)
\end{tabular}\\[3ex]
\begin{tabular}{c|cccccc}
       Pt.& \pointAa{}& \pointAb{}& \pointAc{}& \pointB{}& \pointD{}& \pointDb{} \\ \hline
       $g'$ &1520(330)  &1520(330) &1520(390) &1520(420) &1520(430) &1520(290) \\  \hline
$\Sigma_C$ &1520(100) &1520(100) &1520(100) &1520(10) &1520(50) &1520(50) \\ \hline
$\tilde{g}$&1500(290) &1500(290) &1500(360) &1500(210) &1500(380) &1500(230)
\end{tabular}

\caption{Masses (and widths) in GeV of the lowest color charged
  resonances at tree level in the third generation hypermultiplets and
  SU(3) gauge multiplets. Only decays to two parton final states are
  considered here.\label{tabspectrum}}
\end{center}

\end{table}

\section{Production of the LSP}
\label{sect_lhc}
\subsection{Heavy Quarks and Missing Energy at the LHC}
Stable neutralinos (and very likely any other DM candidates which are
not yet ruled out by direct searches) are invisible to the detectors
at collider experiments. This means that, even though our LSP might be
produced copiously in processes of the type $q\overline{q}\rightarrow
\chi^0 \chi^0$, the remainders of the protons go down the beam line,
and since we lack information about the total cross section, the
process is entirely invisible.  Also, we can not use the recoil
spectrum in $q\overline{q}\rightarrow \chi^0 \chi^0 j$ because the
longitudinal momenta of partons are not known.

The most promising approach is therefore to observe the associated
production with other easily detectable objects such as leptons or
hard jets, and to look for missing momentum in the direction
transverse to the proton beams, $\ptmiss$, which experiments are
equipped to measure by balancing the visible momentum. One such
process which we will however not pursue further is the electroweak
production channel of neutralinos and charginos by VBF.
Instead we concentrate on another set of
final states which is particularly favored in the model which we
consider in this work, the production of third generation quarks in
association with missing energy. In general, it can be an important
process for discovering scenarios of WIMP DM. It is particularly
interesting when a ``partner'' of heavy quarks exists, which decays
directly to a heavy quark (preferably the top) and invisible particles
such as in SUSY with R parity. In our scenario, these partners are the
first stop modes $\tilde{t}$ and they are in a convenient mass range:
still light enough to be pair produced copiously at the LHC at 14 TeV,
and at the same time heavy enough ($m_{\tilde{t}}-m_{\chi}-m_t\gtrsim
400$ GeV) in all but the extreme cases to produce large missing energy
contributions from the decay. Such a situation has been discussed in a
generic way in \cite{Han:2008gy}.  The analysis carried out by the
authors is valid for our $\tilde{t}$ pair production contributions,
but this is only one of the contributions to this class of final
states in our model, where the production of heavy quark and gluon
resonances proves to be important as well. Due to the large number of
interaction vertices in the model and the corresponding number of Feynman
diagrams, we simulate only four parton final states, which
means that we do not consider the possible decay modes of the~$t$.
When judging the results for the
missing energy signal with $t$ and $b$ quarks in the final state, one
therefore has to remember the following points: the $t$ pair
production itself does not introduce $\ptmiss$, but the leptonic and
semileptonic decay modes contain neutrinos, and considering the
relative strength of $t$ pair production, this can constitute an
important background. In addition to this, there are SM processes with
have final states distinguishable from our signal only by their
kinematics, for example $pp\rightarrow b\overline{b}\nu l jj$
\cite{Han:2008gy} or $pp\rightarrow b\overline{b}\nu l \nu l$
corresponding to semileptonic and leptonic decays of top pair
production. While a complete analysis of these backgrounds is beyond
the scope of this work, there is a generic way to suppress them: as
will become apparent in the following discussion, we can
afford to place rather strong $\pt$ and $\Delta \phi(q,\bar q)$ cuts without
losing too much of our signal - this strategy would be futile if there
was only a small mass gap between the $t$, neutralinos and the heavy
partners. We demonstrate this for the leading SM background from
leptonic top decays.

\begin{figure}
 \begin{center}
  \begin{minipage}{0.3\linewidth}
\begin{center}
\includegraphics[width=1.0\textwidth]{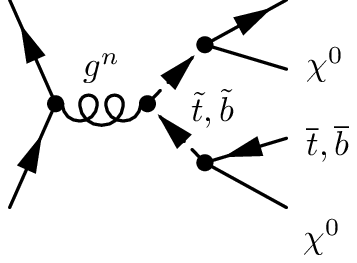}
\end{center}
\end{minipage}
  \begin{minipage}{0.3\linewidth}
\begin{center}
\includegraphics[width=1.0\textwidth]{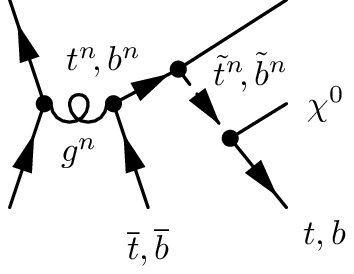}
\end{center}
\end{minipage}\\[1ex]
  \begin{minipage}{0.3\linewidth}
\begin{center}
\includegraphics[width=1.0\textwidth]{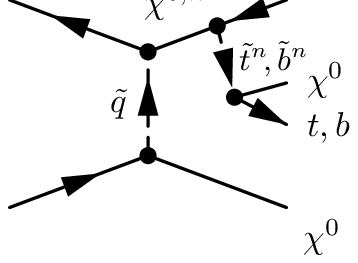}
\end{center}
\end{minipage}
  \begin{minipage}{0.3\linewidth}
\begin{center}
\includegraphics[width=1.0\textwidth]{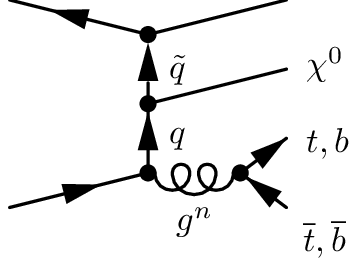}
\end{center}
\end{minipage}\\[1ex]
 \end{center}
\caption{The relevant tree level contributions (up to charge conjugation and crossing) to associated heavy quark and LSP pair production with a $q\overline{q}$ initial state.\label{fig_qqprod}}
\end{figure}

Let us have a closer look at the model to see how the production of
heavy quarks is suited for the discovery of our DM candidate via
$\ptmiss$.  The couplings of the neutralino to light matter are of
electroweak strength, and considering the vertices in the model, the
only production process at $\mathcal{O}(\alpha_s\alpha)$ is through
squarks. However, even though these are Kaluza-Klein in our version of
UV brane symmetry breaking, they can be produced on shell and decay
through \begin{equation}\tilde{q}\longrightarrow
  q\chi^{0\pm}\end{equation} due to the heavyness of the gluino,
making this effectively an $\mathcal{O}(\alpha_s)$ process which is
however suppressed by the large parton momentum fraction $x$ needed to
produce two onshell squarks. We can also potentially obtain final
states with two quarks and two LSPs by producing a heavy chargino or
neutralino resonance which then decays to the LSP and two quarks
through the decay chain shown in figure~\ref{fig_qqprod},
\begin{eqnarray}
 q\overline{q}\rightarrow \chi^{0} \chi^{0\pm,2/3},& & \chi^{0\pm,2/3}\rightarrow \tilde{q}\overline{q} \rightarrow  \chi^0 q\overline{q}+c.c\\
 q\overline{q}\rightarrow \chi^{0} \tilde{g},& & \tilde{g}\rightarrow \tilde{q}\overline{q} \rightarrow  \chi^0 q\overline{q} +c.c.
\end{eqnarray}
but these channels are suppressed because of the small coupling of a light neutralino to squarks and light quarks and, like for LSP annihilation, they do not play a significant role. Kaluza-Klein top production and decay (figure~\ref{fig_qqprod}) is not as suppressed because of the IR localization (and wavefunction deformation from IR masses) of the top quark, and we can thus have
\begin{eqnarray}
 q\overline{q}\rightarrow \overline{t} t^{2}& & t^{2}\rightarrow \chi^0 \tilde{t}^1 \rightarrow \chi^0 \chi^0 t
\end{eqnarray}
as unique new contributions in addition to pure squark pair
production. For example at point \pointAa{} demanding $\pt(q_i)>300\mbox{
  GeV}$, the KK top at $970$ GeV contributes $\Delta\sigma \approx
3.5$ fb.

\begin{figure}
\begin{center}
   \begin{minipage}{0.3\linewidth}
\begin{center}
\includegraphics[width=1.0\textwidth]{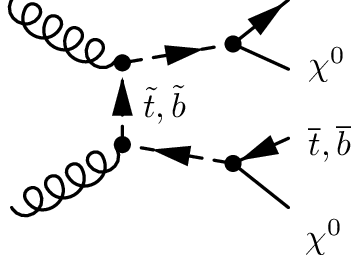}
\end{center}
  \end{minipage}
   \begin{minipage}{0.3\linewidth}
\begin{center}
\includegraphics[width=1.0\textwidth]{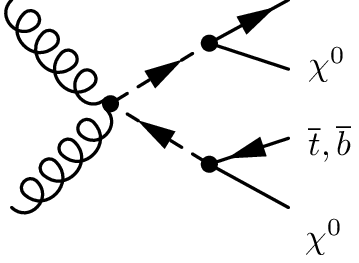}
\end{center}
  \end{minipage}
    \begin{minipage}{0.3\linewidth}
\begin{center}
\includegraphics[width=1.0\textwidth]{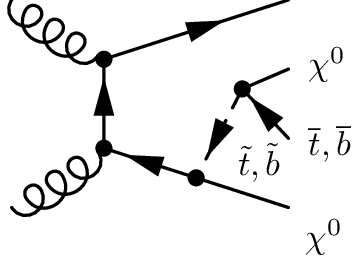}
\end{center}
   \end{minipage}\\[1ex]
   \begin{minipage}{0.3\linewidth}
\begin{center}
\includegraphics[width=1.0\textwidth]{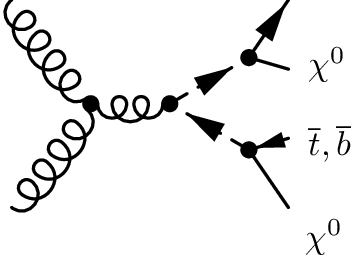}
\end{center}
  \end{minipage}
  \begin{minipage}{0.3\linewidth}
\begin{center}
\includegraphics[width=1.0\textwidth]{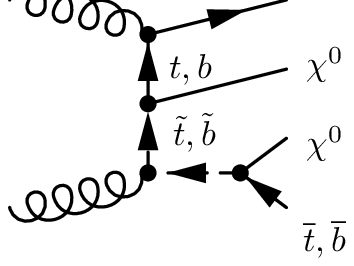}
\end{center}
\end{minipage}
   \begin{minipage}{0.3\linewidth}
 \begin{center}
\includegraphics[width=1.0\textwidth]{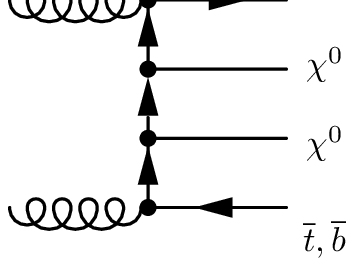}
 \end{center}
 \end{minipage}\\[1ex]
\end{center}
\caption{The relevant tree level contributions (up to charge conjugation and crossing) to associated heavy quark and LSP pair production with a $gg$ initial state.\label{fig_ggprod}}
\end{figure}

We simulate the production of our LSP with WHIZARD, and all following
statements refer to hadronic cross sections using the {MSTW2008} LO
PDF sets from the LHAPDF library \cite{Whalley:2005nh}.  The SM cross
section for $pp\rightarrow b\overline{b},t\overline{t}$ at
$\sqrt{s}=14\mbox{ TeV}$ is very large compared to our signal, at
around $\sigma > 10^4 \mbox{ fb}$ for cuts $\pt(q_i)>300\mbox{
  GeV}$. One should avoid as much as possible misattributions of such
processes, for example by cutting on the angular distance between the jets. At
the $2\rightarrow 4$ parton level, the SM background from
$pp\rightarrow t \overline{t} \nu \overline{\nu}$ is important, and we
want to separate it from our contributions with massive
neutralinos. One efficient way to do this is by cutting on the
transverse momentum of the quarks. For $\pt(q)>300\mbox{ GeV}$ the
cross section is reduced down to $\approx 10 \mbox{ fb}$. The
remaining neutrinos are dominantly produced from $Z$
decays. 

\subsection{Production of $t\overline{t}$ + $\etmiss$ from Neutralinos}
\begin{figure}
\begin{center}
\hspace{-3ex}\includegraphics[width=7cm]{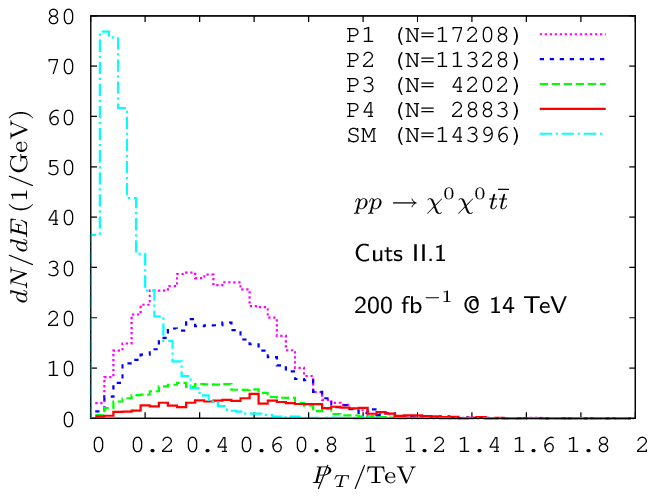}\hspace{-1ex}\includegraphics[width=7cm]{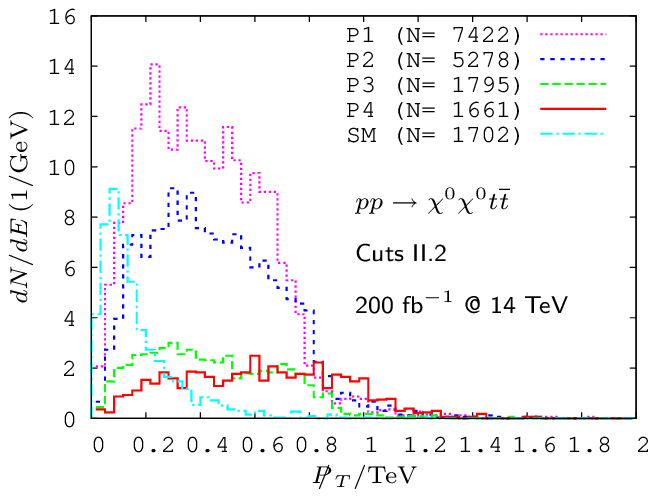}\\
\hspace{-3ex}\includegraphics[width=7cm]{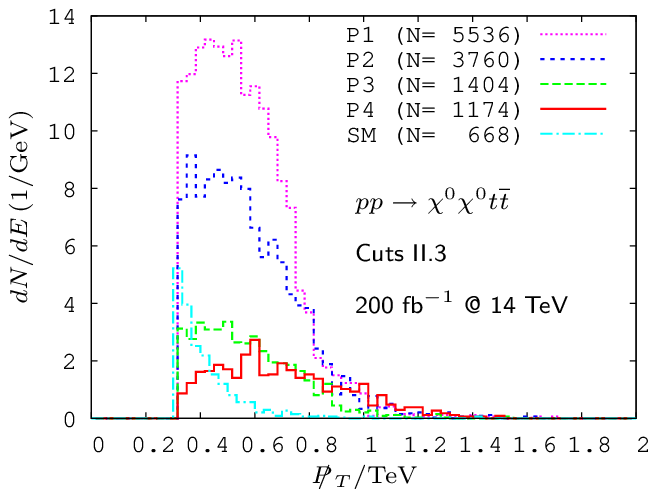}\hspace{-1ex}\includegraphics[width=7cm]{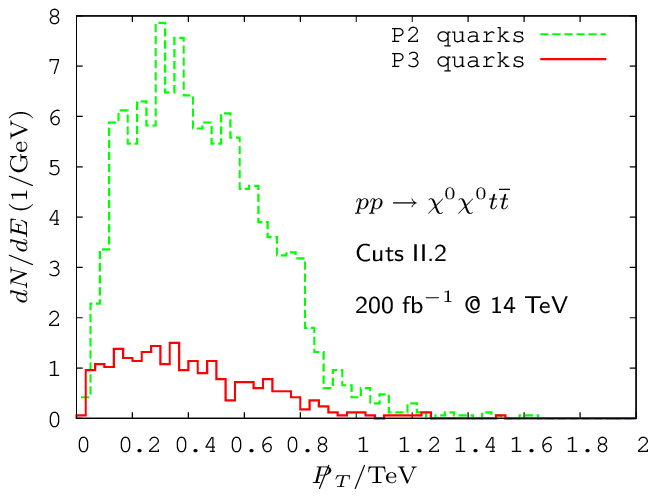}
\end{center}
\caption{Missing energy from LSP and neutrino production in
  association with top pairs for different quark localizations and
  cuts on invariant masses and azimuthal angle (all
  MSTW08). The line marked SM
  shows the missing energy in $pp\rightarrow \nu \overline{\nu} t
  \overline{t}$. For one set of cuts the results for quarks in the
  initial state are compared for two different
  localizations.\label{missingptsummary}}
\end{figure}

\begin{table}
\begin{center}
\begin{tabular}{c|c|c|c|c|c}
Variable&I &II.1 & II.2&II.3&II.3' \\
\hline\hline $\pt(q)$,$\pt(\overline{q})$ &-&$>100$ GeV&$>300$ GeV&$>100$ GeV&$>100$ GeV\\
\hline $ \pt(q,\overline{q})$ &-& -& -& $>300$ GeV& $>300$ GeV\\
\hline $\Delta\phi(q,\overline{q})$ &-&-&-&$[0,160^\circ]$&-
\end{tabular}
\end{center}
\caption{Definitions of kinematic cuts on visible final state
  partons. The $q$, $\overline{q}$ refer to the two quarks in the
  four parton final state, $\Delta\phi$ denotes the azimuthal
  angular distance in the lab frame.\label{tabcuts}}
\end{table}
The SM background $pp\rightarrow \nu\overline{\nu}t\overline{t}$ is
only moderately large. The classes of diagrams contributing to the
signal at this order are shown in figure~\ref{fig_qqprod} for initial
quarks and in figure~\ref{fig_ggprod} for initial gluons.  In addition to
the standard cuts $M(q,\overline{q})\in[10,\infty]$;
$M(\mbox{parton},q),M(\mbox{parton},\overline{q})\in[-\infty,-10]$;
$E(\mbox{parton})>20$; $\eta(q),\eta(\overline{q}) \in [-5,5]$ we
introduce the set of cuts shown in table~\ref{tabcuts} to further
suppress backgrounds. The cuts on the transverse momentum of the
quarks, II.1 and II.2, are efficient in reducing the $2\rightarrow 4$
parton SM background $pp\rightarrow \nu \overline{\nu} t
\overline{t}$.  Rather than imposing stronger bounds on the transverse
energy of the quarks, we can use the azimuthal distance distribution.
We introduce a further set of cuts to exploit the larger contribution
of top quarks from LSP production which are not back to back, II.3.
Using these cuts, we vary the quark bulk masses according to
table~\ref{tab:Pn}, and simulate events corresponding to an integrated
luminosity of $\int\mathcal{L}=200\mbox{ fb}^{-1}$ at $\sqrt{s}=14$
TeV.  The results for the final state $\chi^0\chi^0 t \overline{t}$
are shown in figure~\ref{missingptsummary}. While the masses of the
lightest $\tilde{t}$ states have the largest influence on the total cross
section, the branching ratio of $\tilde{t}$ to LSPs determines how
strong the signal will be for this final state alone. The impact of
contributions from KK gluons is illustrated by the comparison between
points \pointAa{} and \pointAb{}, the latter of which has vanishing
lefthanded $qqg'$ couplings. We find that cross sections can be
reduced dramatically for exact delocalization. The remaining
contributions are basically due to the production of onshell stop
pairs which entails a typical $\pt$ distribution with a cutoff at
twice the $\chi$ or $t$ momentum given by the kinematical function,
\begin{equation}
\pt(q\overline{q}) < \lambda^{1/2}(m_{\tilde{t}}^2,m_t^2,m_\chi^2)/m_{\tilde{t}}
\end{equation}
which is approximately equal to $m_{\tilde{t}}$ for $m_{\tilde{t}}\gg m_{t} \approx m_\chi$.

\subsubsection{Production of $b\overline{b}$ + $\etmiss$ from Neutralinos}
Essentially the same statements are true for this final state, except
that the SM background $pp\rightarrow \overline{\nu}\nu \overline{b}b$
is larger (in particular in the low $\ptmiss$ region), and the signal
cross section is smaller due to the large branching fraction of
$\Gamma(\tilde{b}\rightarrow t \chi^-)$.  Alternatively, one could use
cuts I and only consider the region $\ptmiss >500$ GeV, but the
resulting distribution has an unremarkable shape which is hard to
distinguish from neutrino production. Thus, this channel alone is not
well suited for missing energy observations.

\subsection{Production and Decay of the NLSP}
\begin{figure}
\begin{center}
\hspace{-3ex}\includegraphics[width=7cm]{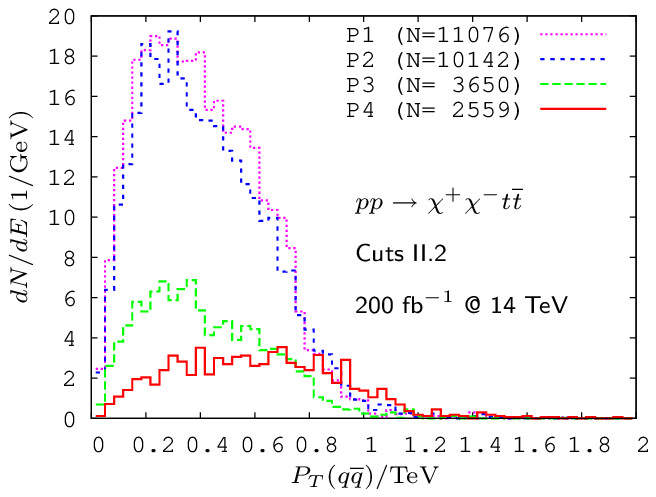}\hspace{-1ex}\includegraphics[width=7cm]{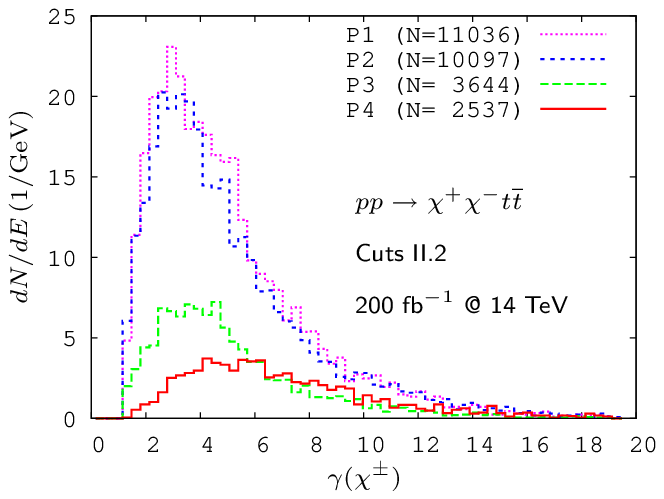}\\
\hspace{-3ex}\includegraphics[width=7cm]{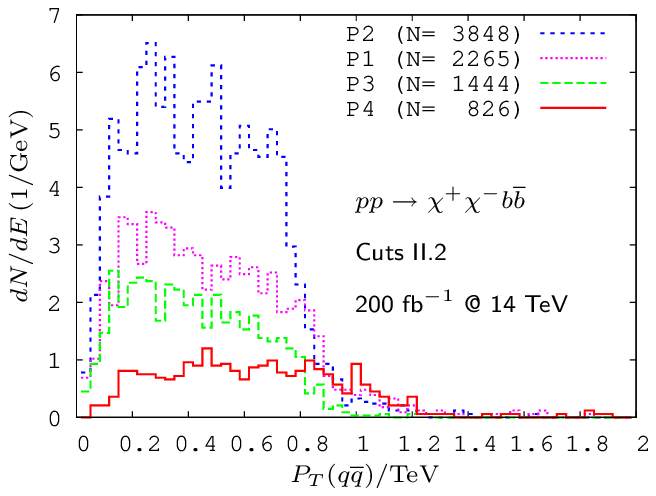}\hspace{-1ex}\includegraphics[width=7cm]{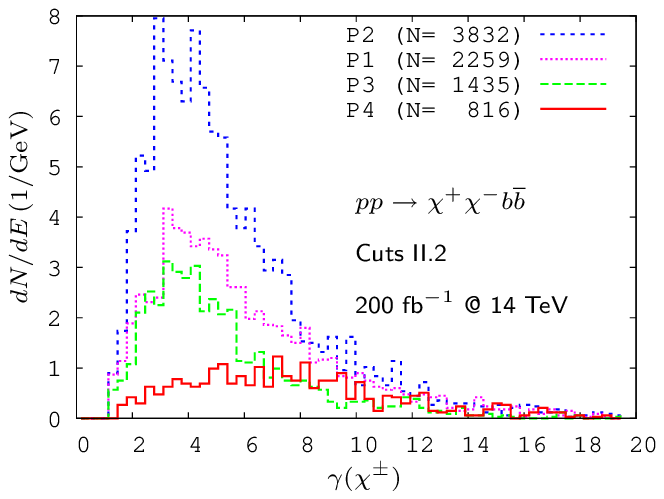}
\end{center}
\caption{Total transverse momenta and boosts of charginos produced in
  association with top and bottom pairs for different quark
  localizations after cuts II.2 (all MSTW08). The total $\pt$ is shown as an
  approximation to $\ptmiss$ which will have further contributions
  from the decay products.\label{fig_chapt}} 
\end{figure}

\begin{figure}
\begin{center}
\hspace{-3ex}\includegraphics[width=7cm]{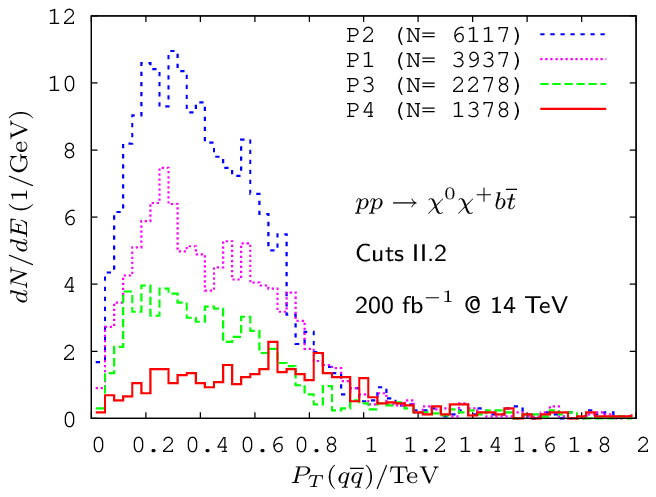}\hspace{-1ex}\includegraphics[width=7cm]{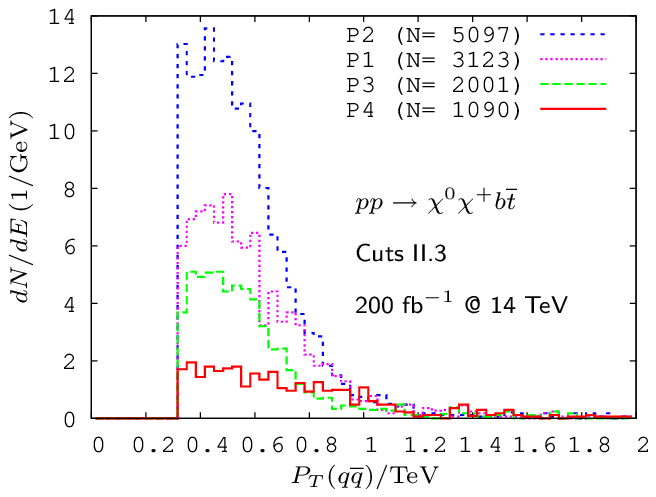}
\end{center}
\caption{Total transverse momenta of electroweak gauginos produced in
  association with $b\overline{t}$ for different quark localizations
  after cuts II.2 and II.3 (all MSTW08). The total $\pt$ is shown as
  an approximation to $\ptmiss$ which will have further contributions
  from the decay products.\label{fig_mixpt}} 
\end{figure}

\begin{figure}
\begin{center}
\hspace{-3ex}\includegraphics[width=7cm]{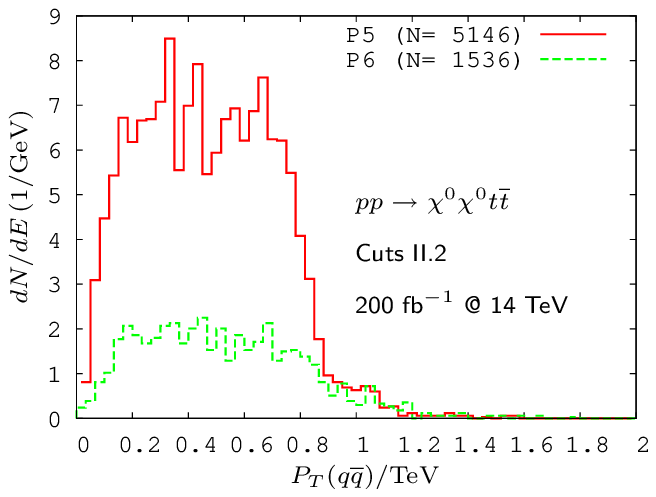}\hspace{-1ex}\includegraphics[width=7cm]{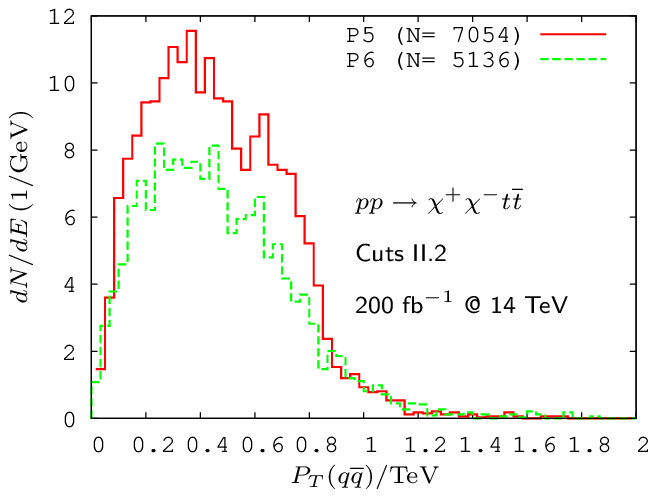}\\
\hspace{-3ex}\includegraphics[width=7cm]{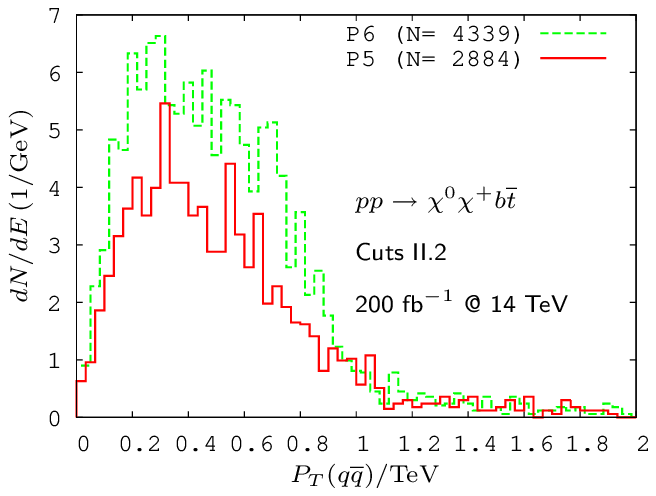}\hspace{-1ex}\includegraphics[width=7cm]{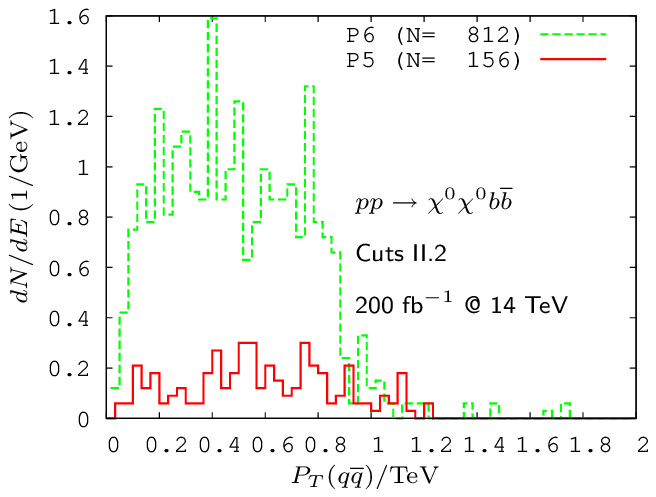}
\end{center}
\caption{Missing energy from LSP and NLSP production for different
  third generation bulk masses. The results from \pointD{} and
  \pointDb{} are compared using cuts II.2. (all
  MSTW08)\label{fig_asymcomp}}
\end{figure}

In the previous sections we have considered the leading order
contributions to neutralino pair production in association with third
generation quarks at the $2\rightarrow 4$ parton level. There is
however another kind of events which might not be easily
distinguishable from LSP production in an experiment, namely the
production of chargino NLSPs or of one chargino NLSP and a neutralino
LSP. Also, there are regions in parameter space
where $$\Gamma(\tilde{q}\rightarrow \chi^\pm
q)\gg\Gamma(\tilde{q}\rightarrow \chi^0 q) $$ for the light third
generation squarks (in particular $\tilde{b}$), making NLSP production
and decay the dominant source of missing energy. Our charginos are
close to the lower experimental exclusion bound and decay via
$\chi^+\rightarrow \chi^0 W^{+*}\rightarrow \chi^0 f_u
\overline{f}_d$. They are very narrow ($\Gamma \approx 10^{-7}\mbox{
  GeV}$) due to the offshell intermediate $W$, yet unlike in some
models with light gravitino LSP, the corresponding lifetime is still
negligible. The fermion pair has an energy of $\approx 10\dots
20\mbox{ GeV}$ in the chargino center of mass frame. Depending on how
well such processes can be resolved in the experiments, this will be
an important channel for the observation of LSP or NLSP production. In
the detector, one might see missing energy with one lepton or
comparatively soft jets (few tens of GeVs) and two $b$ or $t$ jets, so
chargino production events would look similar to the signal discussed
above. We therefore close this section by studying the processes
\begin{equation}
p p \rightarrow \chi^+ \chi^- t \overline t,\chi^+ \chi^- b \overline b,\quad p p \rightarrow \chi^0 \chi^- t \overline b,\quad pp \rightarrow \chi^0 \chi^+ \overline{t} b
\end{equation}
to which the three point couplings $W^{+,n}\chi^-\chi^0$ and $Z^n
\chi^+ \chi^-$ and for the latter also $WWZ$ can make important
contributions, in contrast to neutralino pair production. Since the
results below are for vanishing IR kinetic terms (and thus very heavy
$W$ and $Z$ resonances), these contributions can be both enhanced due
to phase space and decreased further above the energy scale at which
unitarization sets in.  In the case of final states with charginos,
the $\pt$ of the $b$ and $t$ jets is not exactly equal to $\ptmiss$
any more, but only up to the $\pt$ of the softer jet(s) from the
chargino decay products (assuming $t,b$ decay into visible particles
only). The distribution of $\pt(q\overline{q})$ and chargino boosts in
the lab frame is shown in figure~\ref{fig_chapt} for chargino pairs and in
figure~\ref{fig_mixpt} for mixed chargino and neutralino production. The
boost distribution of the charginos can be used to infer how hard the
jets or leptons from the chargino decay products are going to be. The
maximum is at around $\gamma(\chi^\pm)=E_\chi/m_\chi\approx 3\dots 5$
depending on the parameters, leading to rather soft jets, but a
significant number of events can reach $\gamma(\chi^\pm)>5$.  These
cross sections are even larger than those for pure neutralino
production, but it is subject of further study how well this type of
events can be distinguished at the LHC.  The proper treatment of the
background to this type of processes would require production of 8
partons in the final state to account for the three body decays of
two $\chi^\pm$s via offshell $W$s, and possibly detector simulations
which are beyond the scope of this work. In figure~\ref{fig_asymcomp}, we
compare the various final states for two different sets of third
generation localization in which the bulk masses of the $SU(2)_L$ and
$SU(2)_R$ charged hypermultiplets are exchanged. In any case, NLSP
production gives us an equally important source of missing energy.

\subsection{Backgrounds from Leptonic t Decays}

\begin{figure}
\begin{center}
\hspace{-3ex}\includegraphics[width=7cm]{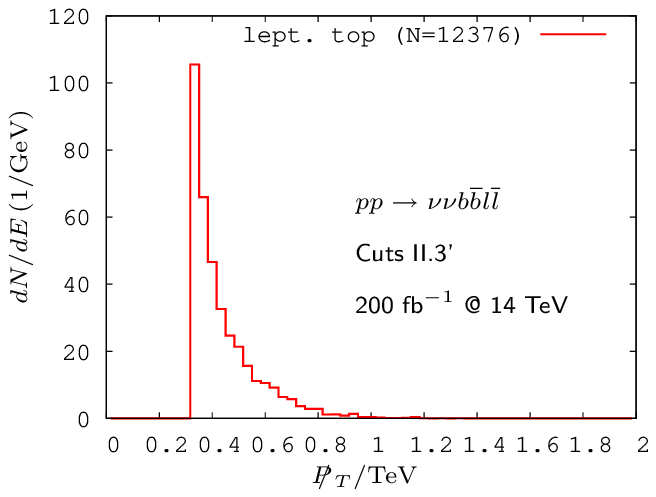}\hspace{-1ex}\includegraphics[width=7cm]{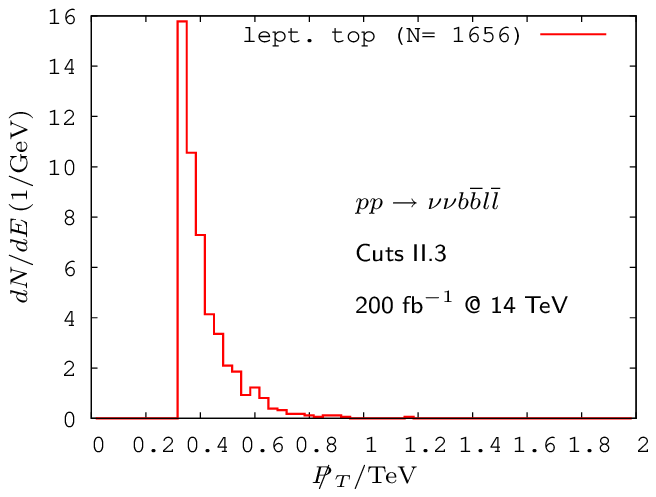}
\end{center}
\caption{Missing energy in $pp\rightarrow \nu \nu b \overline{b} l
  \overline{l}$ (predominantly) from leptonic top decays in the SM
  (all MSTW08). The modified cuts II.3' are II.3 without the cut on
  the azimuthal angular distance to illustrate its
  efficiency.\label{missingptleptonic}} 
\end{figure}

In the detector, the final states $\chi^0 \chi^0 b \overline{b}
q\overline{q} l\overline{\nu}$, $\chi^0 \chi^0 b \overline{b}
q\overline{q} \nu\overline{l} $ or $\chi^0 \chi^0 b \overline{b}
l\overline{\nu}\nu\overline{l} $ as they appear in the processes
discussed above, if one or both tops decay leptonically, and $
b\overline{b}q \overline{q}l \overline{\nu} $, $ b\overline{b}q
\overline{q}\nu \overline{l} $ or $ b\overline{b} l \overline{\nu}
\nu\overline{l} $ (e.g. from top pair production) 
are not distinguishable except for their kinematics, and
all produce missing energy signatures. Both groups can arise from top
decays and thus show the typical edge in the $\overline{l}b$ or
$\overline{b}l$ invariant mass, but while the angle in the center of
mass between the tops is close to 180${}^\circ$ in the case of pair
production, the neutralinos in the signal process may produce substantial
recoil. One might hope that cuts on the azimuthal angle such as cuts
II.3 favor the signal contribution. To check this we implement the
cuts II.2 and II.3 for the visible decay products,
$\Delta\phi(t,\overline{t}) \rightarrow \Delta\phi(bl^+,
\overline{b}l^-)$ and $p_T(t)\rightarrow p_T(bl^+)$ in the SM process
$pp\rightarrow b\overline{b} l^+ l^- \nu \overline{\nu}$. In addition,
we introduce a separation cut $\sqrt{\Delta \phi^2 + \Delta
  \eta^2}>0.3$ for the two $b$ quarks which reduces the SM background
from initial quarks by half. The results are shown in
figure~\ref{missingptleptonic}. Apparently, the angular cut in combination
with $\pt$ cuts is indeed efficient, reducing the missing energy
background from leptonic decays down to $\approx 8 \mbox{
  fb}$. However, when comparing the signal contributions to this, one
has to consider that only a fraction of the top pairs decay purely
leptonically.

\section{Conclusions}
\label{sect_conclusions}
We have given a short overview over our supersymmetric warped model
and an application to the EWSB sector of higgsless models as
introduced in \cite{Knochel:2008ks}. Motivated by the presence of a
realistic CDM candidate and naturally light third generation squark
resonances in this model, we have simulated the production of the LSP
and NLSP in association with third generation quarks with respect to
their missing energy signatures using an implementation in
WHIZARD. Indeed, the production of the lightest $\tilde{t}$ and
$\tilde{b}$ resonances is the most important factor in this class of
processes at the $2\rightarrow 4$ parton level.  Consequently their
mass, which depends on the third generation hypermultiplet bulk mass
parameters $c_{L3}$ and $c_{R3}$, determines the order of magnitude of
cross sections. Since the NLSPs are very light in our model, the final
states involving $\chi^0$ and $\chi^\pm$ are equally important, the
relative strength in each case being determined by the branching
ratios of $\tilde{b}$ and $\tilde{t}$ decays. While the decay products
of the NLSP should in principle be observable as jets or leptons with
energies below $\approx 100$ GeV depending on the mass splitting
$\Delta m(\chi)$, we have used for simplicity the transverse momentum
of the NLSPs as an approximation to $\ptmiss$.

Taking into consideration the leading SM background at the
$2\rightarrow 4$ parton level, $pp\rightarrow \overline{\nu}\nu
\overline{q}q$, we find that cuts on the transverse momenta of final
state quarks are an efficient means of increasing the signal to
background ratio. Depending on the particular decay mode of the tops one is
interested in, missing energy from the resulting neutrinos can become the dominant
SM background. We have illustrated this with a short analysis of the
purely leptonic decay mode of SM top pair production, and have
demonstrated how this contribution can be reduced using $\pt$ and
angular cuts. Again, this background drops quickly at large $\ptmiss$,
and allows us to observe LSP production in the region $\ptmiss\gtrsim
400$ GeV.

The strength of the signal strongly depends on the quark localization
parameters both in the first and third generation. For the following
results we have assumed the cuts II.2, restricting the quark
transverse momenta to $\pt(q)>300$ GeV. For our benchmark points
\pointAc{}, \pointAa{} featuring one rather light $\tilde{b}$ and
$\tilde{t}$ resonance, we find a signal cross section of $\sigma_{s}\approx
130$ fb for the combination of all final states involving
$\chi^{0\pm}\chi^{0\mp}$ at $14$ TeV. The two points yield essentially
the same total cross section and differ only in the relative
importance of the various final states. It drops to
$\sigma_s\approx 60$ fb at $14$ TeV for \pointAb{} with delocalized
light quarks. At \pointB{}, which features IR localized third
generation hypermultiplets, the KK mode splitting is smaller leading
to heavy squarks. The resulting cross section including all final
states is $\sigma_s\approx 40$ fb at $14$ TeV. Worse prospects are
imaginable for more extreme IR localization in combination with
delocalized light quarks.

The production of neutrinos in association with third generation quark
pairs contributes a background cross section of $\sigma_b\approx 20$
fb at $14$ TeV, but mainly at missing momenta $\ptmiss <300$ GeV which
is significantly below the cutoff of $\sigma_s$ at $\pt\approx
\lambda^{1/2}(m_{\tilde{q}}^2,m_q^2,m_\chi^2)/m_{\tilde{q}}$ for most
of the parameter space. In our discussion we have not considered more
complex backgrounds, detector effects or misattributions of final
states in detail.  However we argue that one can apply the discussions
from the literature on $\tilde{t}$ and $\tilde{b}$
production~(e.\,g.~\cite{Han:2008gy}) to our case, keeping in mind
that additional contributions from the KK sector in warped models can
enhance signal cross sections significantly.

The production of heavy intermediates with $m\gtrsim 1$ TeV suffers
severely from reduced parton densities in the large $x$ region at
center of mass energies below $\sqrt{s}=10$ TeV as they are scheduled
for the first run of the LHC. For example, after cuts II.2 and
\pointAa{}, the cross section $\sigma(pp\rightarrow \chi^0\chi^0 t
\overline{t})$ drops by a factor of $\approx 4$.  Fortunately, the
same is true for SM neutrino production, thanks to the strong $\pt$
cut.

\section*{Acknowledgments}
We thank Thomas Trefzger for useful discussions on the ATLAS detector.
This research is supported by Deutsche Forschungsgemeinschaft through
the Research Training Group 1147 \textit{Theoretical Astrophysics and
  Particle Physics}, and by Bundesministerium f\"ur Bildung und
Forschung Germany, grants 05HT6WWA and 05H09WWE.  A.\,K. is also supported by
Deutsche Forschungsgemeinschaft through grant~RE\,2850/1-1.



\begin{thebibliography}{10}


\bibitem{Knochel:2008ks}
  A.~Knochel and T.~Ohl,
  Phys.\ Rev.\  D {\bf 78}, 045016 (2008)
  [arXiv:0805.1379 [hep-ph]].

\bibitem{WHIZARD}
  M.~Moretti, T.~Ohl and J.~Reuter,
  [arXiv:hep-ph/0102195];
  W.~Kilian,
  LC-TOOL-2001-039;
  W.~Kilian, T.~Ohl and J.~Reuter,
  [arXiv:0708.4233 [hep-ph]].

\bibitem{Csaki/etal}
  C.~Csaki, C.~Grojean, L.~Pilo, and J.~Terning,
  \newblock Phys.{} Rev.{} Lett.{} \textbf{92}, 101802 (2004), [arXiv:hep-ph/0308038];
  C.~Csaki, C.~Grojean, H.~Murayama, L.~Pilo, and J.~Terning,
  \newblock Phys.{} Rev.{} \textbf{D69}, 055006 (2004), [arXiv:hep-ph/0305237];
  C.~Csaki, C.~Grojean, J.~Hubisz, Y.~Shirman, and J.~Terning,
  \newblock Phys.{} Rev.{} \textbf{D70}, 015012 (2004), [arXiv:hep-ph/0310355];
  G.~Cacciapaglia, C.~Csaki, C.~Grojean, and J.~Terning,
  \newblock Phys.{} Rev.{} \textbf{D70}, 075014 (2004), [arXiv:hep-ph/0401160];
  G.~Cacciapaglia, C.~Csaki, C.~Grojean, and J.~Terning,
  \newblock Phys.{} Rev.{} \textbf{D71}, 035015 (2005), [arXiv:hep-ph/0409126].

\bibitem{HEIDISUSY}
  N.~Marcus, A.~Sagnotti and W.~Siegel,
  Nucl.\ Phys.\  B {\bf 224}, 159 (1983);
  N.~Arkani-Hamed, T.~Gregoire, and J.~G. Wacker,
  \newblock JHEP \textbf{03}, 055 (2002), [arXiv:hep-th/0101233];
  D.~Marti and A.~Pomarol,
  \newblock Phys.{} Rev.{} \textbf{D64}, 105025 (2001), [arXiv:hep-th/0106256].

\bibitem{Zbb}
  G.~Cacciapaglia, C.~Csaki, G.~Marandella, and J.~Terning,
  \newblock Phys.{} Rev.{} \textbf{D75}, 015003 (2007), [arXiv:hep-ph/0607146].

\bibitem{Whalley:2005nh}
  M.~R.~Whalley, D.~Bourilkov and R.~C.~Group,
  arXiv:hep-ph/0508110.

\bibitem{Han:2008gy}
  T.~Han, R.~Mahbubani, D.~G.~E.~Walker and L.~T.~E.~Wang,
  JHEP {\bf 0905}, 117 (2009)
  [arXiv:0803.3820 [hep-ph]].

\end{thebibliography}
\end{document}